# Probabilistic-Numerical assessment of pyroclastic current hazard at Campi Flegrei and Naples city: Multi-VEI scenarios as a tool for "full-scale" risk management

Giuseppe Mastrolorenzo[1], Danilo M. Palladino[2],*, Lucia Pappalardo[1], Sergio Rossano[3]

[1] *Istituto Nazionale di Geofisica e Vulcanologia, sezione di Napoli, Osservatorio Vesuviano, Via Diocleziano, 328, 80024 Naples, Italy.*

[2] *Dipartimento di Scienze della Terra, Sapienza-Università di Roma, Rome, Italy*

[3] *Dottorato di Ricerca in Dinamica interna dei sistemi vulcanici e rischi idrogeologico-ambientali; Università Federico II, Naples, Italy*

Abstract

The Campi Flegrei volcanic field (Italy) poses very high risk to the highly urbanized Neapolitan area. Eruptive history was dominated by explosive activity producing pyroclastic currents (hereon PDCs; acronym for Pyroclastic Density Currents) ranging in scale from localized base surges to regional flows. Here we apply probabilistic numerical simulation approaches to produce PDC hazard maps, based on a comprehensive spectrum of flow properties and vent locations. These maps are incorporated in a Geographic Information System (GIS) and provide all probable Volcanic Explosivity Index (VEI) scenarios from different source vents in the caldera, relevant for risk management planning. For each VEI scenario, we report the conditional probability for PDCs (i.e., the probability for a given area to be affected by the passage of PDCs in case of a PDC-forming explosive event) and related dynamic pressure. Model results indicate that PDCs from VEI<4 events would be confined within the Campi Flegrei caldera, PDC propagation being impeded by the northern and eastern caldera walls. Conversely, PDCs from VEI 4-5 events could invade a wide area beyond the northern caldera rim, as well as part of the Naples metropolitan area to the east. A major controlling factor of PDC dispersal is represented by the location of the vent area. PDCs from the potentially largest eruption scenarios (analogous to the ~15 ka, VEI 6 Neapolitan Yellow Tuff or even the ~39 ka, VEI 7 Campanian Ignimbrite extreme event) would affect a large part of the Campanian Plain to the north and the city of Naples to the east. Thus, in case of renewal of eruptive activity at Campi Flegrei, up to 3 million people will be potentially exposed to volcanic hazard, pointing out the urgency of an emergency plan. Considering the present level of uncertainty in forecasting the future eruption type, size and location (essentially based on statistical analysis of previous activity), we suggest that appropriate planning measures should face at least the VEI 5 reference scenario (at least 2 occurrences documented in the last 10 ka).

Introduction

Active calderas are among the most hazardous volcanic areas in the world (Lipman, 2000). Caldera volcanism is characterized by rare, large-scale (VEI≥5) eruptions and even super-eruptions (Sparks et al. 2005, Self 2006) and punctuated by more frequent intermediate- (VEI 3-4) or small-scale (VEI 1-2) events. More than one hundred Quaternary calderas worldwide, including the caldera complexes of Rabaul (Papua New Guinea), Yellowstone (USA), Long Valley (USA), Kilauea (USA) and Campi Flegrei (Italy), underwent periods of unrest during the second half of the 20[th] century (Newhall and Dzurisin 1988). The related hazard assessment is complicated by the interactions between the magmatic systems and their volcano-tectonic and hydrogeological/geothermal settings. In particular, the possible roles of the stress field related to the caldera structure and hydrothermal system on the occurrence, location and style of eruptions is still matter of debate (e.g., Gottsman and Martì 2008, Simakin and Ghassemi 2010).

The Campi Flegrei caldera (Fig. 1) poses a volcanic risk ranking among the highest in the world, together with the neighboring Vesuvius volcano (e.g. Orsi et al., 2004, 2009, Rossano et al. 2004, De Natale et al., 2006, Neri et al., 2008, Marzocchi and Woo, 2009, Mastrolorenzo and Pappalardo 2010, Lirer et al., 2010, Chiodini et al., 2012, Pappalardo and Mastrolorenzo, 2012, Selva et al., 2012a, b). A recent study (Selva et al., 2012a), based on a Bayesian Event Tree approach, estimated a monthly probability of eruption at Campi Flegrei of $1.6 \times 10^{-3}$. The extreme risk is due to the possible high explosivity of a future eruption and the very high degree of urbanization of the area, also including the city of Naples (Napoli). Nearly two million people live within 15 km from the centre of the Campi Flegrei caldera (12 km across). The volcanic history of the Campi Flegrei in the last ~50 ka has been dominated by explosive activity featured by intermediate- to large-scale PDCs, monogenetic tuff cone- and tuff ring-forming hydromagmatic events and subordinate Strombolian and Plinian fall events (e.g., Rosi and Sbrana, 1984, Orsi et al., 1996, Mastrolorenzo et al. 2006, 2008). Occasional effusive activity also occurred. The area potentially affected by a future eruption depends primarily on the eruption style, magnitude and source location. The volcanological record shows a full range of possible eruptive magnitudes, mechanisms and vent locations to be considered for probabilistic evaluation.

Hazard scenarios for fallout events have been reported by Mastrolorenzo et al. (2006, 2008) and Costa et al. (2008). Here, we address hazard assessment from PDCs related to the full range of eruptive scenarios at Campi Flegrei. Previous approaches (Orsi et al. 2004, 2009, Lirer et al. 2010, Alberico et al., 2011) were based on the reproduction of the distribution patterns of past PDCs on the present topography, which is somehow misleading; in fact, geomorphic changes over the lifetime of the volcanic field imply that a future event will produce a different PDC distribution with respect to its analogues that occurred in the past. Todesco et al. (2006) and Mele et al. (2015), by numerical simulations of PDC obtained important information on flow propagation, but did not provide hazard maps (e.g. information on the probability in the unit of time) associated with the simulated event. Neri et al. (2015) provided probabilistic PDC invasion maps for a whole-range scenario, based on the last 15 ka eruptive activity. However PDC hazard maps including specific hazard variables (e.g. dynamic pressure) with the related levels of damage for each VEI class are still lacking.

Moreover, inferences on the future eruption type and vent position, based on the extrapolation from the most recent eruptive behavior (i.e., <5 ka; Orsi et al. 2009, Selva et al., 2012b), are highly uncertain. Since at present robust constraints on the future behavior of the Campi Flegrei caldera are lacking, a primary requisite for the development of mitigation and crisis response strategies is to consider a full range of possible scenarios. Even risk mitigation strategies based on elicitation procedures and cost/benefit analyses need volcanological-probabilistic scenarios for each VEI.

In order to assess a comprehensive set of reference eruptive scenarios at Campi Flegrei caldera, we performed numerical simulations of PDCs from explosive events ranging in VEI between 2 and 6, on a new 5 m resolution digital elevation model, to produce probabilistic hazard maps embedded in a GIS framework (Fig. 2-6). This work builds on the model frame of Rossano et al. (2004), which provided the yearly probabilities of occurrence and areal dispersal of PDCs averaged on the whole VEI range, with the qualification that a complete set of distinctive scenarios for each VEI (from 2 to 6) is here considered. The definition of eruption VEI is essentially inferred from the scale of PDC deposits, pyroclastic fall events being quite subordinate at Campi Flegrei. By merging the available field data from past eruptions of different size with a probabilistic approach, we compute the conditional probability of each area to be invaded by PDCs in case of an eruption with a given VEI, relevant for the application of event-tree approach in the management of a volcanic crisis.

Eruptive history of Campi Flegrei

The activity history of the Campi Flegrei volcanic field in the last ~50 ka comprises some large-scale eruptions (e.g. De Vivo et al. 2001, 2006, 2010, Rolandi et al. 2003) and several tens of intermediate- to small-scale eruptions. In particular, the ~39 ka Campanian Ignimbrite super-eruption (VEI 7), with an inferred volume of erupted products in the order of 300 km$^3$ (e.g. Pappalardo et al. 2002, 2008, Pappalardo and Mastrolorenzo 2012), and the ~15 ka Neapolitan Yellow Tuff eruption (VEI 6), with inferred 40 km$^3$ of products, were the dominant caldera-forming events that controlled the volcanic-tectonic evolution of the area.

Between the two largest events and after the Neapolitan Yellow Tuff eruption, the caldera was the site of intense, mostly explosive activity. At least eleven low- to moderate-scale (VEI 2-5) explosive events have been recognized in the stratigraphic sequence between ~39 and 15 ka (Pappalardo et al. 1999). After an eruptive break following the Neapolitan Yellow Tuff eruption, at least seventy explosive eruptions clustered in the last ~10 ka (e.g., Di Vito et al. 1999, D'Antonio et al. 1999, Fedele et al. 2011), ranging in VEI between 2 and 5 and in erupted volume between tens of millions of cubic meters to a few cubic kilometers. These recent events, including the AD 1538 Monte Nuovo eruption, the last one occurred at Campi Flegrei, typically produced monogenetic tuff rings and tuff cones and subordinate spatter cones, scattered throughout the caldera. Commonly, monogenetic centers were formed as a result of different eruptive styles, which produced in turn breccia layers, fallout units, and a variety of deposits from different types of PDCs. Phreatomagmatic activity was largely dominant: both "wet" and "dry pyroclastic surges" from tuff cone- and tuff ring-forming events covered areas of several square kilometers around intracaldera vents (Mastrolorenzo et al., 2001). Subordinately, widespread tephra sheets from Plinian-style fallout and major PDCs, and occasional lava domes and lava flows, were also produced.

The study PDC deposits: implications for modeling

Although field characteristics of PDC deposits at Campi Flegrei are widely described in the literature, detailed data on related eruptive and emplacement mechanisms are reported in relatively few cases (Di Vito et al. 1987, Mastrolorenzo 1994, Wohletz et al. 1995, Rossano et al. 1996, 2004, Dellino et al. 2001, 2004a, 2004b, 2008). In light of detailed model studies (e.g. Dellino et al. 2004a, 2004b, 2008, Rossano et al. 1996, 2004), PDC properties at Campi Flegrei show a wide range of variability. Textural and grain size features of phreatomagmatic PDC deposits indicate an emplacement by dilute to moderately concentrated PDCs (i.e., densities between a few kg/m$^3$ and $10^2$ kg/m$^3$). The latter prevailed in the activity history and were related to several intracaldera tuff cones, as well as to a significant part of the Neapolitan Yellow Tuff major eruption (Orsi et al., 1992, Scarpati et al., 1993). Based on thickness of individual depositional units (ranging between tens of centimeters to a few meters) and the variety of bedforms indicative of bedload transport and sedimentation, flow front depths between a few meters and several tens of meters and relatively low yield strength (<$10^2$ Pa) Bingham rheologies can be inferred, consistent with model data for "pyroclastic surges" worldwide (Sheridan 1979, Brissett and Lajoie 1990, Yamamoto et al. 1993, Sigurdsson et al. 1987, Wilson and Head 1981, Freundt and Schmincke 1986).

Conversely, other PDCs at Campi Flegrei, including small-scale scoria flows (e.g., from the AD 1538 Monte Nuovo eruption) and locally distributed proximal spatter- and lithic-rich units of the Campanian Ignimbrite, show evidence for high-concentration PDCs. From a Bayesian inverse approach considering PDC runout and response to topography (Rossano et al. 1996, 2004), consistent with calculations from clast grading patterns, densities up to $10^3$ kg/m$^3$, viscosities up to $10^3$ Pa s, and high yield strength ($10^2$-$10^3$ Pa) Bingham rheologies are derived, as typical of "pyroclastic flows" *s.s.* (Sheridan 1979, Beget and Limke 1988, 1989, Sparks 1976, Wilson and Head 1981, Yamamoto et al. 1993, Palladino and Valentine 1995).

PDC distribution was controlled to variable extents by local geomorphic features (e.g., caldera walls, intracaldera plains and ridges, cones and craters; Fig. 1), depending on eruption size and vent location. Generally, the propagation of PDCs from low VEI scenarios was strongly controlled by low relief topography (not exceeding a few hundreds of meters in elevation) and, particularly, was impeded by the Posillipo and Camaldoli hills. Instead, major PDCs from VEI>4 eruptions overtopped intracaldera reliefs and even the 400 m-high caldera walls, travelling tens of km over the surrounding plains (de Vita et al. 1999).

Previous work recognized the dominant phreatomagmatic signature of PDC events at Campi Flegrei (Mastrolorenzo 1994, Wohletz et al. 1995, Mastrolorenzo et al. 2001, Dellino et al. 2004a, 2004b, Mastrolorenzo and Pappalardo 2006). In many cases, deposit textures and pyroclast shapes indicate that explosive magma-water interaction was superimposed on magmatic activity and took place after advanced levels of magma vesiculation. The depth and efficiency of magma-water interaction varied from eruption to eruption and even in the course of individual events, over a wide range of eruption intensities from small tuff-cone-forming (e.g., VEI 2 AD 1538 Monte Nuovo) to Phreatoplinian (VEI 6 Neapolitan Yellow Tuff; Scarpati et al. 1993,

Orsi et al. 1993, 1995) events. In these cases, external water drastically perturbed the ascending magma and enhanced its fragmentation (Mastrolorenzo et al. 2001, Mastrolorenzo and Pappalardo 2006). Consequent changes in temperature, grain-size distribution, density and ascent velocity of the erupting mixture eventually controlled the eruptive style and emplacement mechanisms. Thus, the rapid conversion of thermal to mechanical energy due to explosive magma-water interaction may allow even low-scale eruptions to produce relatively high-mobility PDCs, thus enhancing the related hazard. According to the computation of Mastin (1995), phreatomagmatic blasts may achieve high initial pressures and initial velocities (up to 400 m/s). Enhanced magma fragmentation, pressure and exit velocity may drastically change the properties of the erupting gas-mixture and consequent transport and depositional mechanisms (e.g., Wohletz 1983, Wohletz et al. 1995, Valentine 1987, Dellino et al. 2004a). Generally, tuff-ring-forming events at Campi Flegrei mostly produced relatively hot, dry, dilute PDCs ("dry surges"), while predominant tuff-cone-forming events produced PDCs with lower temperature and higher particle concentration ("wet surges"), which were more effectively controlled by topography (de Gennaro et al. 1998). Notably, these contrasting characteristics are independent on the eruption scale and may refer to the full VEI 2-6 range, including the largest phreatomagmatic event, the Neapolitan Yellow Tuff.

Conversely, PDCs from purely magmatic, Plinian-style events are quite subordinate in the Campi Flegrei activity record. Thus, the application of numerical simulations of PDCs derived from eruption column collapse (e.g., Todesco et al. 2006) is not suitable for the large majority of PDCs at Campi Flegrei. At the extreme scale, the Campanian Ignimbrite caldera-forming eruption (and older analogue events) produced regional density-stratified PDCs, as well as locally dispersed, topography-controlled, spatter- and lithic-rich concentrated PDCs in proximal settings (i.e.,

Piperno and Breccia Museo units) and pumice-rich concentrated PDCs in distal ones (Fisher et al. 1993, Rosi et al. 1996).

Owing to the large uncertainties in the type and size of the future event at Campi Flegrei, in this work we adopt a field-based approach to address a full spectrum of potential VEI scenarios. To constrain reference data for PDC-forming events at Campi Flegrei to be adopted in numerical simulations, we have conducted field investigations on representative PDC deposits. Table 1 summarizes the data for representative PDC events selected from eruptions with different VEI. Of note, even for equivalent VEI, changes in volcanic and geomorphic contexts (i.e., vent location, initial conditions at PDC generation, substrate topography) may generate largely different PDC behaviors (transport mode, velocity, dynamic pressure, temperature, etc.) and resulting dispersal patterns.

Table 1 - Reference eruptions of Campi Flegrei

| Pyroclastic Formation | Age (Ka) | Volcanological classification | Volcanic Explosivity Index (VEI) | eruptive magnitude | average total volume ($km^3$) | Maximum runout (km) | Inferred initial velocity of PDCs (m/s) | *density ($kg/m^3$) | *thickness (m) | *velocity (m/s) |
|---|---|---|---|---|---|---|---|---|---|---|
| Campanian Ignimbrite | 39 | low aspect ratio ignimbrite | 7 | 7,5 | 150 | > 80 | 160 – 220 | - | - | - |
| Breccia Museo | 39 | block and ash flow | 5 | 5,0 | 2,5 | - | - | - | - | - |
| Neapolitan Yellow Tuff | 14,9 | hydromagmatic flow/surge sheet | 6 | 6,5 | 40 | ca. 30 | 180 – 370 | 12 (20 km) | 320 | 63 |
| Gauro | <12 | tuff cone | 4 | 4,5 | 1,5 | 3,4 | 129 | - | - | - |
| Miseno | 12 to 9.5 | tuff cone | 2 | 2,5 | 0,1 | - | - | - | - | - |
| Nisida | 12 to 9,5 | tuff cone | 2 | 2,0 | 0,02 | - | - | - | - | - |
| Mofete | >10 | tuff cone | 2 | 2,0 | - | - | - | - | - | - |
| Archiaverno | 10,7 | tuff ring | 4 | 4,5 | - | - | - | - | - | - |
| Fondi di Baia | 8,6 | tuff cone | 2 | 1,5 | 0,03 | 1,3 | 80 | - | - | - |
| Baia | 8,6 | tuff ring | 3 | 2,5 | - | 1,4 | 82 | - | - | - |
| Cigliano | 4,5 | cinder cone | 2 | 2,0 | 0,03 | - | 76 | - | - | - |
| Solfatara | 4 | tuff ring | 3 | 3,0 | 0,07 | 2,1 | 101 | - | - | - |
| Agnano Monte Spina | 4,1 | flow/surge sheet | 5 | 5,3 | - | 22 | 328 | 5 (3 km) | 76 | 39 |
| Astroni | 4.1-3.8 | tuff ring | 4 | 4,5 | 1,00 | 3 | 121 | 10 (1 km) | 87 | 34 |
| Averno | 3,9 – 3.7 | tuff ring | 3 | 3,5 | 0,50 | 2,8 | 117 | 7.3 (1.5 km) | 110 | 21 |
| Monte Nuovo | 1538 AD | cinder cone | 2 | 2,5 | 0,04 | 1,3 | 80 | 54 (1 km) | 64 | 38 |

* results of PDCs stratified model for the dilute currents (see text for further explanations); in brackets it is indicated the distance from the vent of the sampling sites for calculations

The present PDC modeling builds on the mass-independent kinematic approach for gravity-driven PDCs (McEwen and Malin 1989; also applied to Campi Flegrei and Somma-Vesuvius; Rossano et al. 1996, 1998, 2004, Mastrolorenzo et al. 2006, 2008), which considers the spreading of a wide category of PDCs (ranging from dilute turbulent blasts to high-density, non-turbulent flows and lahars) on a 3D model of topographic surface. In particular, we follow this model approach aiming at adequately describing the regional features of PDCs (i.e., flow path, distal reaches, topography control) to adopt in hazard assessment, rather than the details of local flow structure and sedimentation.

In dilute gas-particle dispersions, transport is driven by the motion of the "fluid" mixture (e.g., fluid medium + suspended particles) in response to the density contrast with the ambient fluid, and particle interactions are negligible (i.e., true suspension current; *fluid gravity flows*, Hsü 1989). Instead, in highly concentrated gas-particle dispersions, it is the motion of solid particles in response to gravity that makes the interstitial fluid move and particle interactions dominate the transport system (*sediment gravity flows*, Hsü 1989). Referring to the wide spectrum of PDC dynamics (e.g., following the conceptual frame of Valentine 1987, Druitt 1998, Palladino and Simei 2002), we remark that the present modeling is more apt to describe self-sustained, moderate- to high-particle concentration PDCs (i.e., driven by dense, non-turbulent basal avalanches, possibly with associated dilute ash clouds), as well as the concentrated basal portions of thick, density-stratified, turbulent PDCs. Instead, model simplifications and assumptions (see below) limit its applicability to highly dilute, turbulent PDCs (i.e., suspension currents driving bedload motion) or to the dilute upper portions of thick stratified PDCs. For instance, model frame does not account for thickness and density variations that occur downcurrent due to air entrainment and pyroclast deposition, both being major controlling factors of dilute PDC dispersal, together with mass eruption rate at flow inception (Bursik and Woods 1996).

Indeed, other Authors (Todesco et al. 2006, Esposti Ongaro et al. 2008) modeled PDCs in the Neapolitan area essentially as strongly inflated, turbulent clouds resulting from Plinian-type collapsing columns. However, even in this case, model results (Neri et al. 2003) show that high-particle concentration basal portions of turbulent PDCs become increasingly important with increasing grain size of the erupting mixture. Moreover, recently observed eruptions (cf. Druitt 1998 and reference therein) provide evidence that the dense basal part of a stratified turbulent current may detach as a concentrated underflow, outrun significantly the parent current, and spread as far as the most distal reaches. Thus, in PDC modeling, focusing on the concentrated underflow rather than the parent dilute current appears more appropriate to capture the general PDC behavior relevant for hazard issues on a regional scale. Also, we stress that high-particle concentration flows pose locally the highest impact on the anthropic environment, since they exhibit the highest values of the three factors responsible of casualties and damages: dynamic pressure, heat and suffocation capability.

Computational model

The reference eruptions considered in table 1 put constraints on the range of variables that can be used in predictive numerical models for PDC hazard estimates at Campi Flegrei. In order to perform quantitative assessment of the PDC hazard related to each VEI scenario, we adopt the volcanological-probabilistic approach in Rossano et al. (1996, 1998, 2004), Mastrolorenzo et al. (2006) and De Natale et al. (2006), where PDCs are modeled on the basis of a simple gravity-driven model (Malin and Sheridan 1982, Sheridan 1979, Sheridan and Malin 1983, Wohletz and Sheridan 1979).

*PDC physical model*

The physical model adopted in our numerical simulations of PDCs is an improved version from Rossano et al. (1996, 2004; see model details therein). The approximation of the gas-pyroclasts system to a continuum shearing flow can be suitable to describe the overall behavior of moderate- to high-density PDCs on a macro-scale and has been also used to explain some relevant aspects of PDC deposits (e.g., vertical and lateral coarse-tail grading, Palladino and Valentine 1995). We recall the set of equations that describe the motion of Bingham and Newtonian fluids in an infinitely wide channel (McEwen and Malin 1989). The steady, uniform vertical velocity profile is (see also table 2 for definition of notations):

$$v(z) = \frac{1}{\eta}\left[\frac{\rho g \sin\theta (D^2 - z^2)}{2} - k(D-z)\right] \quad (1)$$

where $z \geq D_c$ is the height (measured from the bottom of the channel), k is the yield strength (equal to zero for a Newtonian fluid), $\rho$ is the flow density, g is the acceleration due to gravity, $\theta$ is the ground slope, $\eta$ is the flow viscosity, D the total flow depth, and $D_c$ is the plug thickness:

$$D_C = \frac{2(kD + \eta v) - \sqrt{(2kD + 2\eta v)^2 - 4k^2 D^2}}{2k} \quad (2)$$

The acceleration of the plug is:

$$\frac{dv_p}{dt} = a - \frac{2k}{\rho(D + D_c)} - \frac{2\eta v_p}{\rho(D^2 + D_C^2)} \quad (3)$$

where $v_p$ is the plug velocity and a is the component of the acceleration due to gravity along the flow direction, which also takes into account ground friction and turbulence resistance.

Flow motion is described by the mean cross-sectional velocity:

$$v = \frac{\int_{D_c}^{D} v(z) dz + v_p D_c}{D} \quad (4)$$

The resistance terms in the Bingham flow equation depend on several factors. The transition from laminar to turbulent regime in a Bingham flow depends upon two dimensionless numbers: the Reynolds number, Re = $\rho$ v D/$\eta$, and the Bingham number, Bi = k D/$\eta$ v. From empirical relations (Middleton and Southard 1978), when Bi exceeds about 1.0, the onset of turbulence occurs for Re/Bi ≥ 1000. Following McEwen and Malin (1989), the frontal air drag is neglected, so that the deceleration of the entire flow is due to air drag on its upper surface:

$$\frac{dv}{dt} = -c_a \left(\frac{\rho_a}{\rho}\right) \frac{v^2}{2D} \qquad (5)$$

where $\rho_a$ is the air density, and the drag coefficient for atmosphere, ranges between 0.1 and 1 (Perla 1980). Since flow deceleration is proportional to $\rho_a/\rho$ air drag will not be significant for relatively dense flows.

Our model describes flow motion as a family of trajectories of invidual 1D flow fronts, generated radially from each vent, moving on a 3D model of topographic surface with given kinematic and rheological properties. Since the PDC-ambient density contrast is not considered by the model, the distal reaches of the flow (i.e., when velocity drops to zero) essentially depend on the total energy balance of the moving flow, including conservative and dissipative energies.

**Table 2.** Definition of notations through the text.

| Symbol | Definition |
|---|---|
| A | component of the acceleration due to gravity along the flow |
| Bi | Bingham number |
| $c_a$ | Drag coefficient for atmosphere |
| D | Flow thickness |
| $D_c$ | Plug thickness |
| dv/dt | Acceleration of the flow |
| G | acceleration due to gravity |
| K | Yield strenght |
| Re | Reynolds number |
| T | Time |
| V | Mean cross-sectional velocity |
| $v_0$ | Initial velocity of the flow |
| $v_p$ | Velocity of the plug |
| Z | Distance within flow measured from ground |
| η | Viscosity |
| θ | Slope of ground |
| ρ | Density of flow material |
| $\rho_a$ | Density of atmosphere |

*Input data: eruptive vents*

In spite of a huge improvement of knowledge on volcano-tectonic setting, magma evolution and eruptive history of Campi Flegrei (e.g., Pappalardo et al. 2002, 2012; Fedele et al. 2011, Cannatelli 2012, and references therein), the vent location of a future eruption remains highly uncertain. The activity younger than 15 ka shows a possible relationship between vent locations and tectonic lineaments (e.g., Miseno-Baia and Concola-Minopoli, D'Antonio et al. 1999), as small-scale events with mafic magma compositions tend to concentrate near caldera margins, while the majority of eruptions with felsic compositions tend to occur from volcano-tectonic structures scattered throughout the caldera. However, in the last 5 ka, although eruptions were often localized in the central part of the caldera, within ~2 km from the present town of Pozzuoli (e.g., Solfatara, Agnano-Montespina, and Astroni eruptions), a number of eruptions (e.g., Archiaverno, Averno, Nisida, Baia, Miseno and the youngest event of AD 1538 Monte Nuovo) were scattered over a wider area extending up to the caldera rims, thus indicating that a future event might occur anywhere in the caldera, including its margins. On the other hand, there is no evidence of an eruptive source outside the caldera rim in the last 15 ka. Moreover, Isaia et al. (2009) reported the contemporaneous eruptions from vents located in different sectors of the caldera during the volcanic activity occurred in the last 4.1 ka.

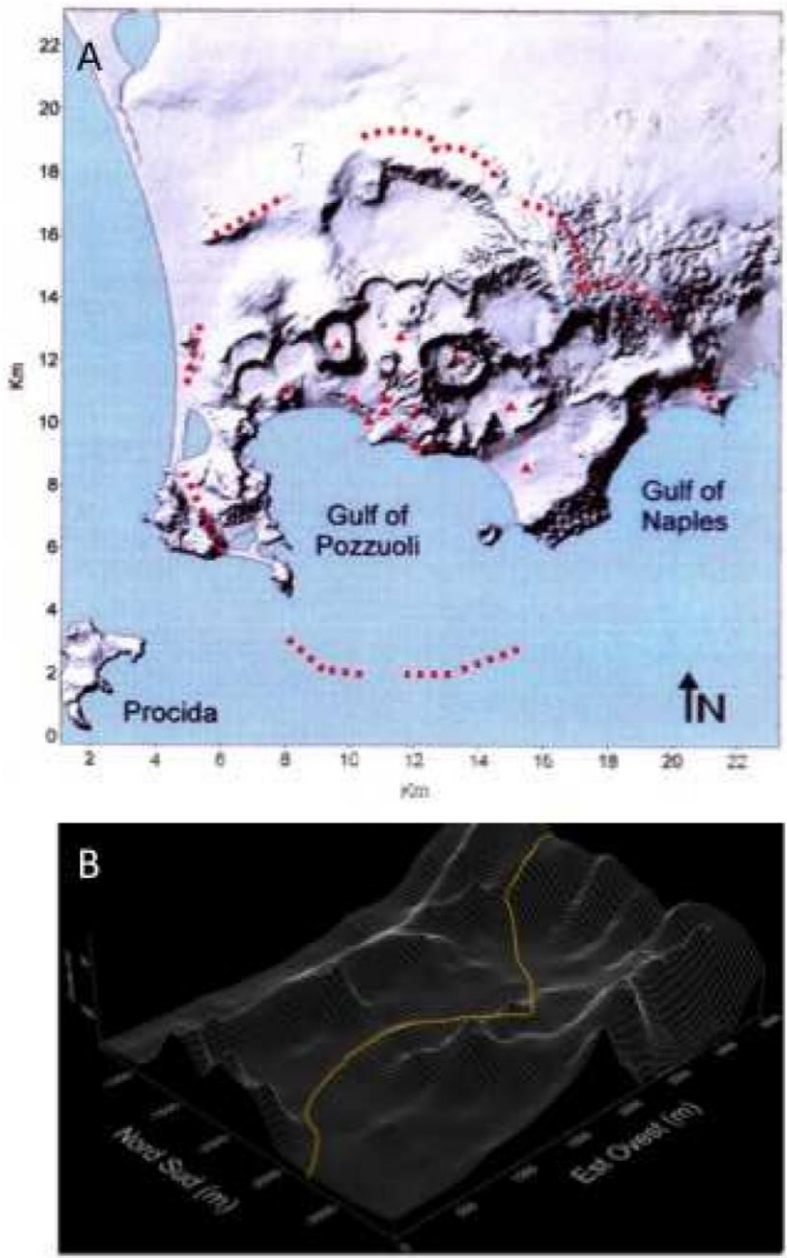

Figure 1. *A. Sketch shadow relief map of the Campi Flegrei volcanic field. Vent locations adopted for numerical simulations of PDCs at Campi Flegrei are reported: fourteen possible eruptive vents are considered, i.e.: six located within the source area of the most recent eruptions, the other ones in the area with the highest horizontal (four vents) and vertical (four vents) deformation during recent bradyseismic events (see text for further explanation). B. Example of the motion of a material point along a digitalized 3D surface considered in the present model (see "Computational model" section for explanation).*

On these grounds, following Rossano et al. (2004), in order to explore a representative set of vent locations (including central, intermediate and peripheral zones of the caldera, as well as intra-caldera plains and reliefs), we have fixed a set of fourteen vents with homogeneous level of probability, regularly spaced along three concentric arcs centered at Pozzuoli (Fig. 1A). This is considered the centre of the Campi Flegrei caldera, based on the youngest (<5 ka) eruption cluster, the peak of bradyseismic and fumarolic activities in the last few decades, and the pattern of gravimetric anomalies.

From each vent, numerical modeling simulated a family of flow trajectories generated in all directions on a 3D model of topographic surface, considering a specific set of PDC properties for each VEI scenario (see following). Each flow path and corresponding distal reach thus results from a specific combination of input PDC parameters and interaction with topography.

*Input data: PDC properties for the different VEI scenarios*

On the grounds of the previous model frame (Rossano et al., 1996, 2004), and following the approach of Mastrolorenzo and Pappalardo (2010) for Somma-Vesuvius, here we simulate the reference scenarios for the PDCs at Campi Flegrei, potentially associated with eruptions of different VEI. In particular, we have created matrixes of input data for eruptions with VEI ranging from 2 to 6, including the values of physical parameters and their range of variation inferred from the studied events (table 3), i.e.: flow velocity at the vent, flow thickness and rheological parameters of gas-particle mixtures (density, viscosity, and yield strength). In addition, we consider the characteristic parameters of VEI 7 events (e.g., the 39 ka Campanian Ignimbrite and possible analogous examples documented in the early activity record, De Vivo et al.

Table 3 – Input parameters for PDC simulations

|  | dilute PDCs | | | | | concentrated PDCs | | | |
| --- | --- | --- | --- | --- | --- | --- | --- | --- | --- |
|  | VEI2 | VEI3 | VEI4 | VEI5 | VEI6 | VEI2 | VEI3 | VEI4 | VEI5 |
| height (m) | 5÷10 | 5÷20 | 10÷100 | 20÷200 | 30÷300 | 1÷5 | 1÷10 | 2÷20 | 5÷20 |
| density (kg/m3) | 2÷30 | 2÷100 | 2÷100 | 2÷100 | 2÷100 | 200÷1500 | 200÷1500 | 200÷1500 | 200÷1500 |
| viscosity (Pa/s) | 0.00002 | 0.00002 | 0.00002 | 0.00002 | 0.00002 | 1÷2000 | 1÷2000 | 1÷2000 | 1÷2000 |
| initial velocity m/s | 10÷50 | 10÷70 | 10÷200 | 30÷300 | 30÷300 | 5÷20 | 5÷30 | 10÷70 | 10÷100 |
| yield strength Pa | 0 | 0 | 0 | 0 | 0 | 1÷2000 | 1÷2000 | 1÷2000 | 1E-1÷2000 |

To obtain an empirical estimate of the initial flow velocities in the input matrix, preliminary calculations were performed by adopting the "energy line" (or, in three dimensions, the "energy cone") approach for granular flows, based on the maximum runout distances of PDCs actually recognized for each VEI class at Campi Flegrei (table 3), and assuming the typical range of values of the Heim coefficient (0.1-0.8, Sheridan 1979) for events of analogous scale worldwide. A flow viscosity of $2 \times 10^{-5}$ Pas (corresponding to pure hot steam) and yield strength of 0 Pa have been fixed for Newtonian-type, highly dilute PDCs, while a range of values (table 3) has been considered for moderate- to high-particle concentration PDCs (e.g., Mc Even and Malin 1989).

According to literature data (e.g., Sparks 1976, Freundt and Schmincke 1986, Valentine 1987, Palladino and Valentine 1995, Dellino et al. 2008), flow density values may range between a few kg/m³ in the very dilute portions of PDCs and even >2000 kg/m³ (typical of rock slide avalanches) in the basal portions of high-particle concentration PDCs. In order to constrain our simulations to actual PDCs occurred at Campi Flegrei, we consider a wide range of input PDC densities. For concentrated PDCs, input data in table 3 were retrieved from Rossano et al. (1998). For relatively dilute PDCs, density values were obtained from a Matlab code developed by Mastrolorenzo and Pappalardo (2010). Following the calculation method of Dellino et al. (2008), this code computes flow thickness, density and velocity at a given site, based on grain size data (Md ) and pyroclast densities in the related deposit. On these grounds, a density range between 10 and 100 kg/m³ has been investigated for modeling low- to moderate-concentration PDCs. Notably, the flow velocities calculated by the code are broadly consistent with those inferred from the energy line approach.

The sampling of the values reported in the matrix of table 3 allow us, for each VEI class, to generate families of numerical PDC trajectories propagating from each vent in all directions on the Campi Flegrei Digital Elevation Model (source: Laboratory of Geomatica e Cartografia, INGV-OV Naples). Comprehensive volcanologic–probabilistic scenarios are obtained by combining the entire set of computer simulations for each VEI .

In particular, the model considers the motion of a material point with rheological properties along a digitalized 3D surface (Fig. 1B), subdivided in triangles that are equilateral of side 250 m in plan view. Site by site, the motion of the point (i.e.

acceleration, deceleration, deviation) depends on the value and direction of the ground slope. From each eruptive vent, 360 individual flows were generated in all radial directions (i.e. a single flow for each degree of the direction). Flow lines deviate from their initial direction and, in places, intersect each other depending on the morphology. The hazard associated to each triangle of territory is thus proportional to the ratio of the number of flows that cross the triangle area vs. the total number of flows generated.

Results

The simulated scenarios, considering a spectrum of PDC concentrations, are shown in Fig. 2-6 for each VEI. Model outputs yield the maximum PDC travel distances in all directions and the number of flow passages for each unit area, i.e. proportional to the probability of a given locality to be affected by the passage of PDCs (computed following Rossano et al. 2004) in the case of a specific VEI event.

As stated above, this model approach can be applied to describe the regional features of PDCs (i.e., flow path, distal reaches, topography control) relevant for hazard assessment, rather than the detailed PDC behavior in terms of local fluid dynamics and sedimentation. Moreover, model outputs provide the local PDC impact at specific localities. In case of occurrence of a given VEI event (i.e., conditional probability equal to one), Fig. 2-6 report for each locality the probability for the passage of relatively dilute to concentrated PDCs and the associated maximum dynamic pressures (calculated following the approach of Valentine 1998). The first assessment of PDC-induced damage to structures and natural environment, corresponding to different values of dynamic pressure, was based on the effects of nuclear explosions (Valentine 1998).

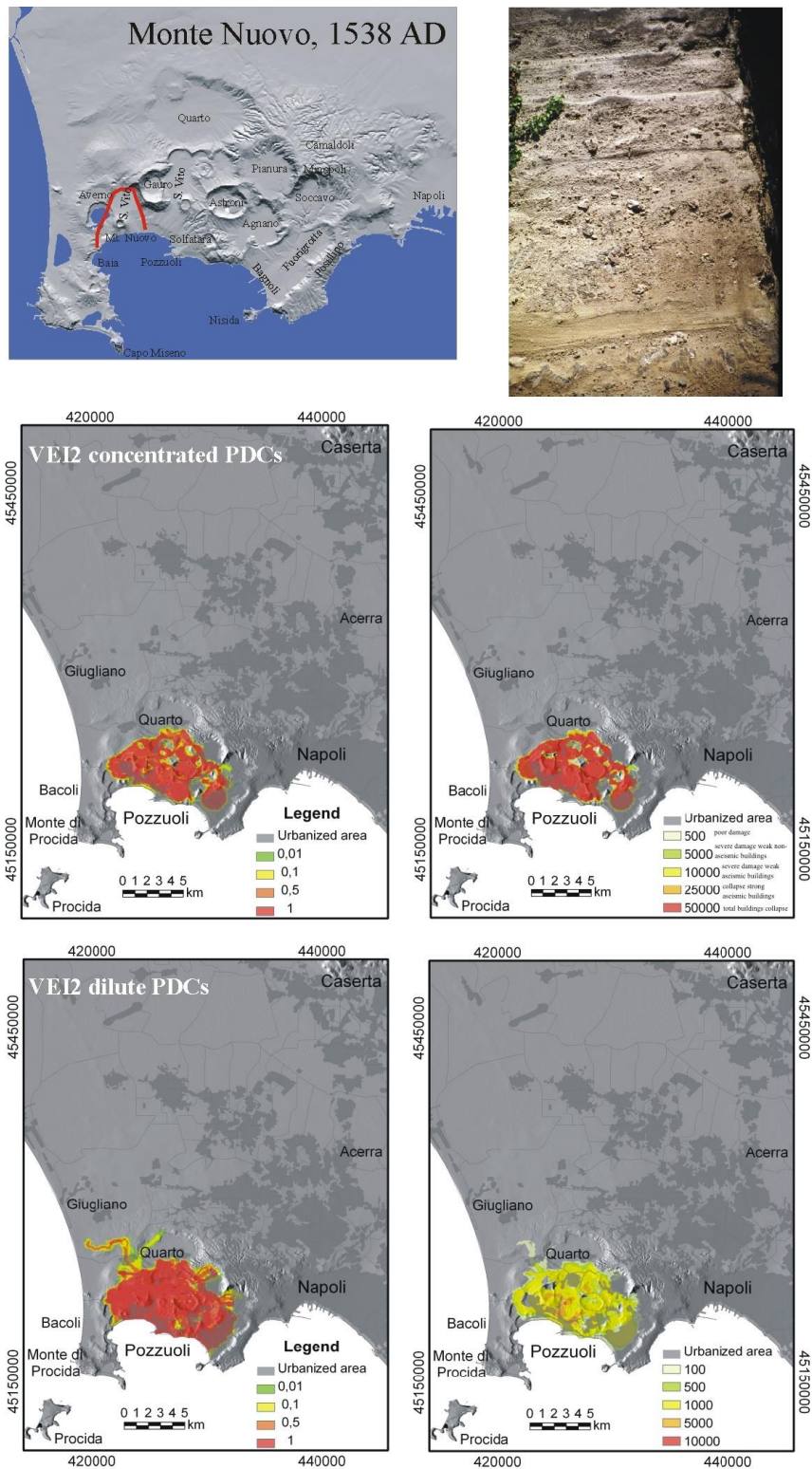

*Figure 2. Model results for VEI 2 eruptions at Campi Flegrei. A) Maximum limit (red line) reached by the PDCs during the 1538 AD Monte Nuovo eruption, representative of VEI 2 events at Campi Flegrei; B) typical exposure of the Monte Nuovo PDC deposits at about 0.5 km from the vent; C) hazard map of conditional probability (i.e., probability of a given point to be affected by the passage of PDCs in the case of an eruption of a given VEI) and D) the associated maximum dynamic pressures (expressed in Pa) for moderate- to high-particle concentration PDCs; E) hazard map of conditional probability and F) the associated maximum dynamic pressures for dilute PDCs (see text and tables 1, 3 for the PDC characteristics adopted in the simulations). Levels of damage associated to different values of dynamic pressure (after Valentine, 1988): 500 Pa= poor damage; 5,000 Pa= window failure, lower limit for severe damage and collapse of weak non-aseismic buildings; 10,000 Pa= limit for severe damage and collapse of weak aseismic buildings; 25,000= limit for collapse of strong aseismic buildings and volcanic masonry walls. The urbanization pattern is also shown in C-F.*

More specifically, experimental results reported for typical buildings of the Neapolitan area (Petrazzuoli and Zuccaro 2004) and Montserrat (Baxter et al. 2005), indicate that: ~1 kPa is the lower threshold value of dynamic pressure for structure damage; severe building damage and collapse may occur for dynamic pressures between 5 and 16 ka; extensive to total building collapse may occur for values in the order of a few tens of kPa.     Our numerical simulations allow us to explore the areal patterns of variably concentrated PDCs, over the likely range of eruption sizes. In order to provide regular zoning of the hazard parameters, we have applied a contouring algorithm based on spline interpolation to the output data. The donut hole around the vents in some maps of Fig. 2-6 is due to the adopted pattern of flow sources, i.e., families of radial 1-D flow lines, starting from a circle with a radius of 250 m centered on each vent.     The hazard maps reported in this work are suitable to be incorporated in a geographic information system (GIS) and made available via the web, in order to render them usable by local and government officials and expert users for risk mitigation and education management

*VEI ≤ 3 scenario*

During VEI 2 (Fig. 2) and VEI 3 (Fig. 3) eruptions, variably concentrated PDCs are usually confined in the caldera and strongly controlled by topography. PDCs can advance only 2 km from the vent; their distribution can be either subcircular or directional, depending on vent location and surrounding topography. The PDC dynamic pressure drops sharply with distance. Nevertheless, in the immediate vicinity of the vent, high-density PDCs can produce very high dynamic pressures (>10 kPa).

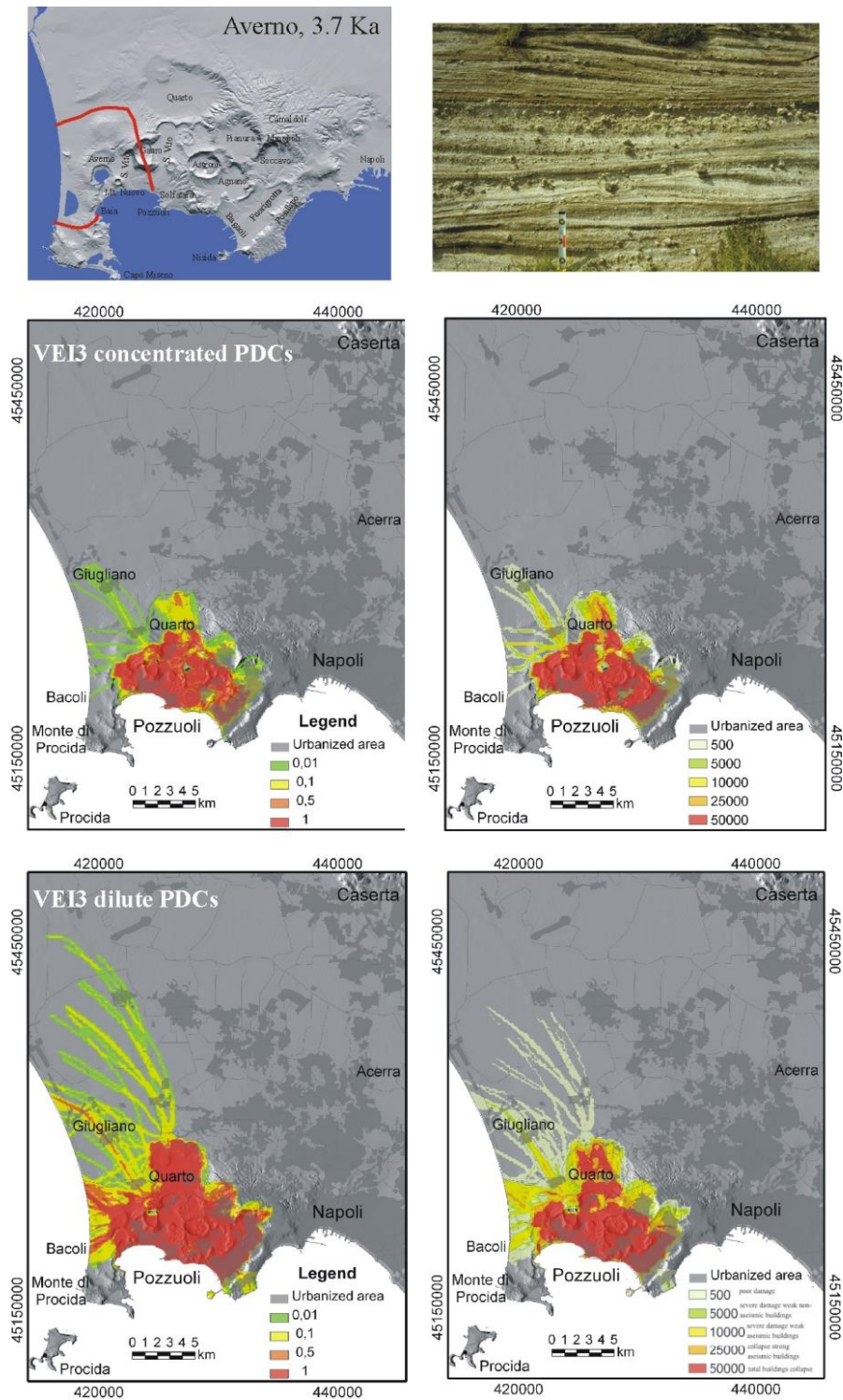

*Figure 3. Model results for VEI 3 eruptions at Campi Flegrei. A) Maximum limit (red line) reached by the PDCs during the ~ 3.9-3.7 ka Averno eruption, as example of VEI 3 events at Campi Flegrei; B) typical exposure of the Averno PDC deposits at about 1.5 km from the vent; C) hazard map of conditional probability and D) the associated maximum dynamic pressures (expressed in Pa) for moderate- to high-particle concentration PDCs; E) hazard map of conditional probability and F) the associated maximum dynamic pressures for dilute PDCs (see text and tables 1, 3 for the PDC characteristics adopted in the simulations). Levels of damage associated to different values of dynamic pressure as in Fig. 2. The urbanization pattern is also shown in C-F.*

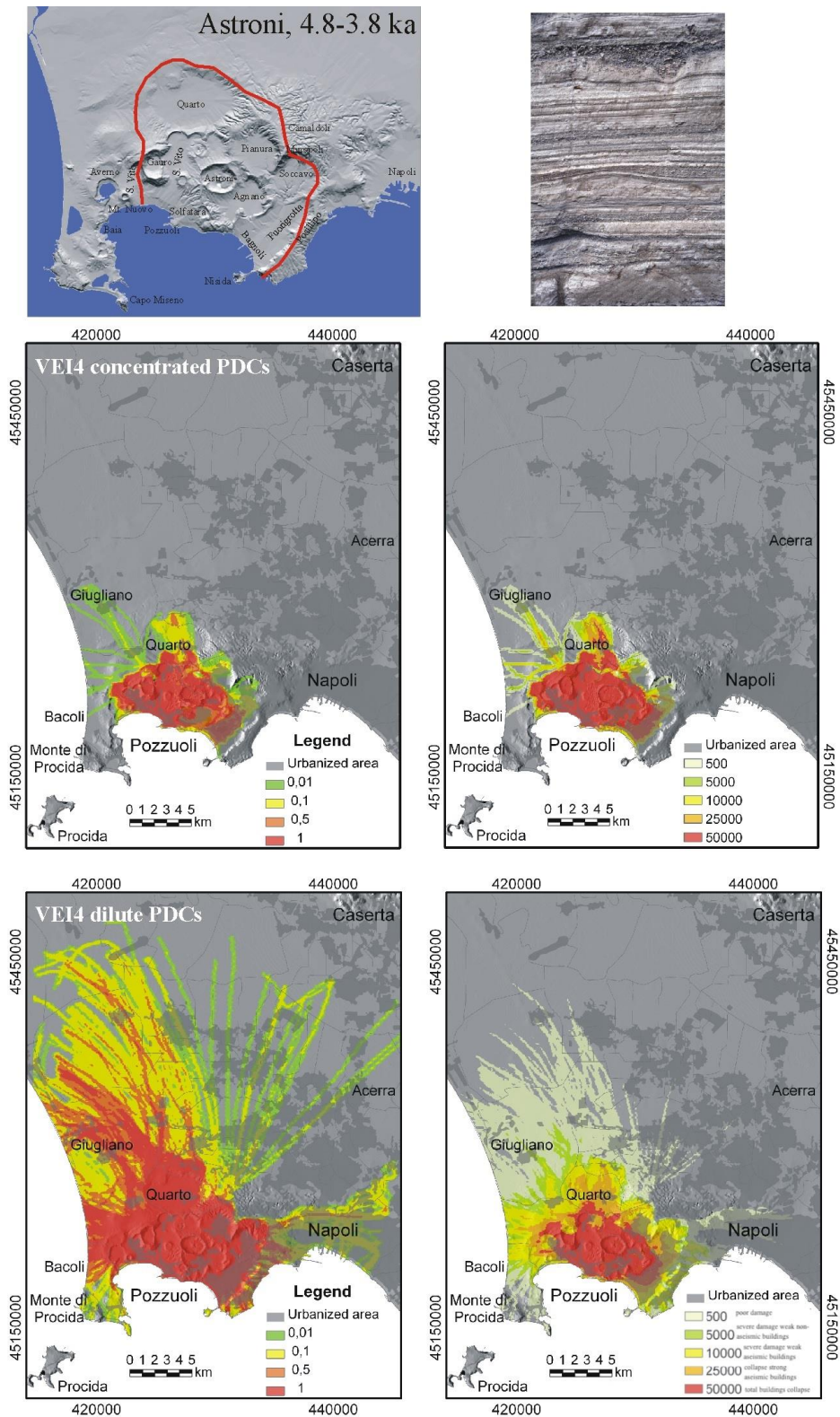

*Figure 4. Model results for VEI 4 eruptions at Campi Flegrei. A) Maximum limit (red line) reached by the PDCs during the ~ 4.1-3.8 ka Astroni eruption, as example of VEI 4 at Campi Flegrei; B) typical exposure of the Astroni PDC deposits at about 1.5 km from the vent; C) hazard map of conditional probability and D) the associated maximum dynamic pressures (expressed in Pa) for moderate- to high-particle concentration PDCs; E) hazard map of conditional probability and F) the associated maximum dynamic pressures for dilute PDCs (see text and tables 1, 3 for the PDC characteristics adopded in the simulations). Levels of damage associated to different values of dynamic pressure as in Fig. 2. The urbanization pattern is also shown in C-F*

*VEI 4 scenario*

In this scenario, PDCs are moderately controlled by topography and propagate as far as 3-5 km from the vent on average, being mostly confined in the caldera (Fig. 4). However, in cases of vents located in valleys, PDCs can advance to distances even exceeding 10 km. The Camaldoli and Posillipo hills always act as major barriers for PDCs of different concentration, so that the area Northeast of the caldera wall is sheltered from PDCs. In particular, relatively dilute, highly mobile PDCs, also capable to preserve high temperature over long distance (Mastrolorenzo et al. 2010), may affect a wider area than high-concentration PDCs, although with lower local dynamic pressures. Overall, VEI 4 PDCs pose very high risk in the whole highly urbanized district of Pozzuoli, as well as in the western suburbs of Naples (e.g., Bagnoli, Fuorigrotta, Posillipo).

*VEI 5 scenario*

PDCs of this VEI class are limitedly controlled by intra-caldera topography and may propagate at distances even exceeding 25 km from the vent, well beyond caldera rims (Fig. 5). However, different from relatively dilute PDCs, high-concentration PDCs are effectively stopped by the ca 400 m high Camaldoli hill barrier, along the northeastern caldera ridge. Due to high mobility and regardless the vent position, PDCs of this class would impact the whole caldera area, including the western suburbs of Naples (e.g., Bagnoli, Fuorigrotta, Posillipo). The capability to impact also the Naples city center depends on the vent position and PDC parameters, and it may result higher for thick, dilute PDCs, capable to overtop topographic barriers and/or develop concentrated underflows at considerable distances from vent.

*VEI ≥6 scenario*

Eruptions of this size have occurred two times in the last 40 ka, i.e., the VEI 6 Neapolitan Yellow Tuff (ca. 15 ka; Fig. 6) and the VEI 7 Campanian Ignimbrite (ca. 39 ka), and possibly several times in the last 300 ka (e.g., De Vivo et al. 2001). PDCs from these VEI scenarios cover a wide range of sizes, up to the most powerful PDCs of the Campanian Ignimbrite dispersed on a regional scale.

Model output for a VEI 6 event (Fig. 6) yields PDC runout distances even in excess of 30 km. Due to the high capability to overpass topographic highs, it appears that wide sectors of the Campanian Plain beyond caldera rims would be affected by the passage of PDCs of this magnitude. Topographic barrier effects become significant only in the most distal areas, as flow velocity decreases. This poses extremely high risk for the whole Campi Flegrei caldera, the Naples district and surrounding areas of the Campanian region.

In addition to the modeled VEI 6 scenario, field distribution of the Campanian Ignimbrite (e.g., Fisher et al., 1993) indicates that PDCs from VEI 7 extreme events may extensively propagate north- and eastward as far as Roccamonfina volcano and intra-Apennine valleys, as well as southward across the sea and eventually overpass the Sorrento Peninsula.

The Campanian Ignimbrite PDCs have been described in terms of a thick, low-concentration, turbulent "regional transport system", feeding a high-concentration, topography-controlled, "local depositional system" (Fisher et al. 1993). This behaviour may produce locally independent concentrated PDCs, resulting in high dynamic pressures even in most distal settings

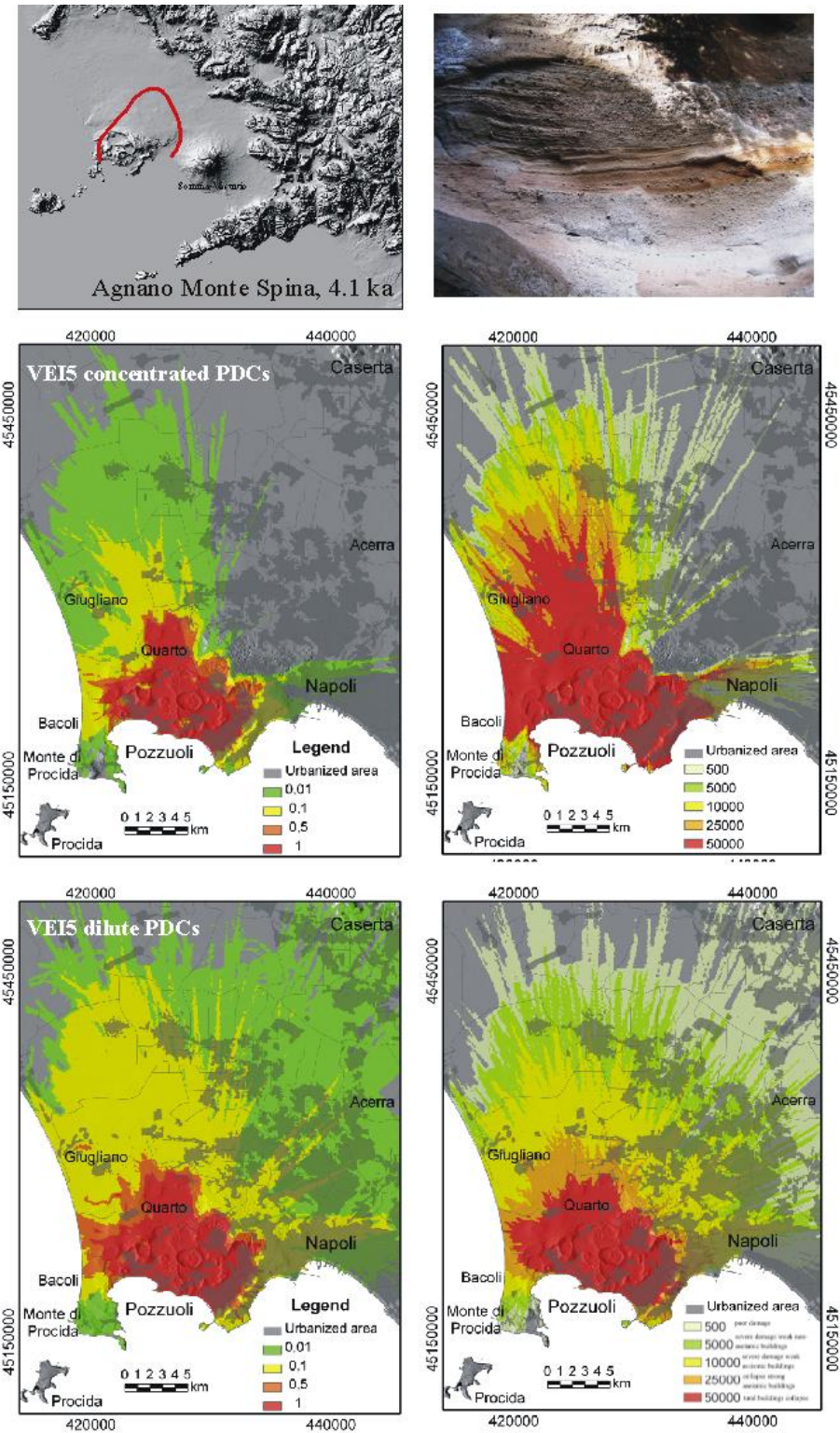

*Figure 5. Model results for VEI 5 eruptions at Campi Flegrei. A) Maximum limit (red line) reached by the PDCs during the ~ 4.1 ka Agnano Monte Spina eruption, as example of VEI 5 at Campi Flegrei; B) typical exposure of the Agnano Monte Spina eruption PDC deposits at about 2 km from the vent; C) hazard map of conditional probability and D) the associated maximum dynamic pressures (expressed in Pa) for moderate- to high-particle concentration PDCs; E) hazard map of conditional probability and F) the associated maximum dynamic pressures for dilute PDCs (see text and tables 1, 3 for the PDC characteristics adopted in the simulations). Levels of damage associated to different values of dynamic pressure as in Fig. 2. The urbanization pattern is also shown in C-F.*

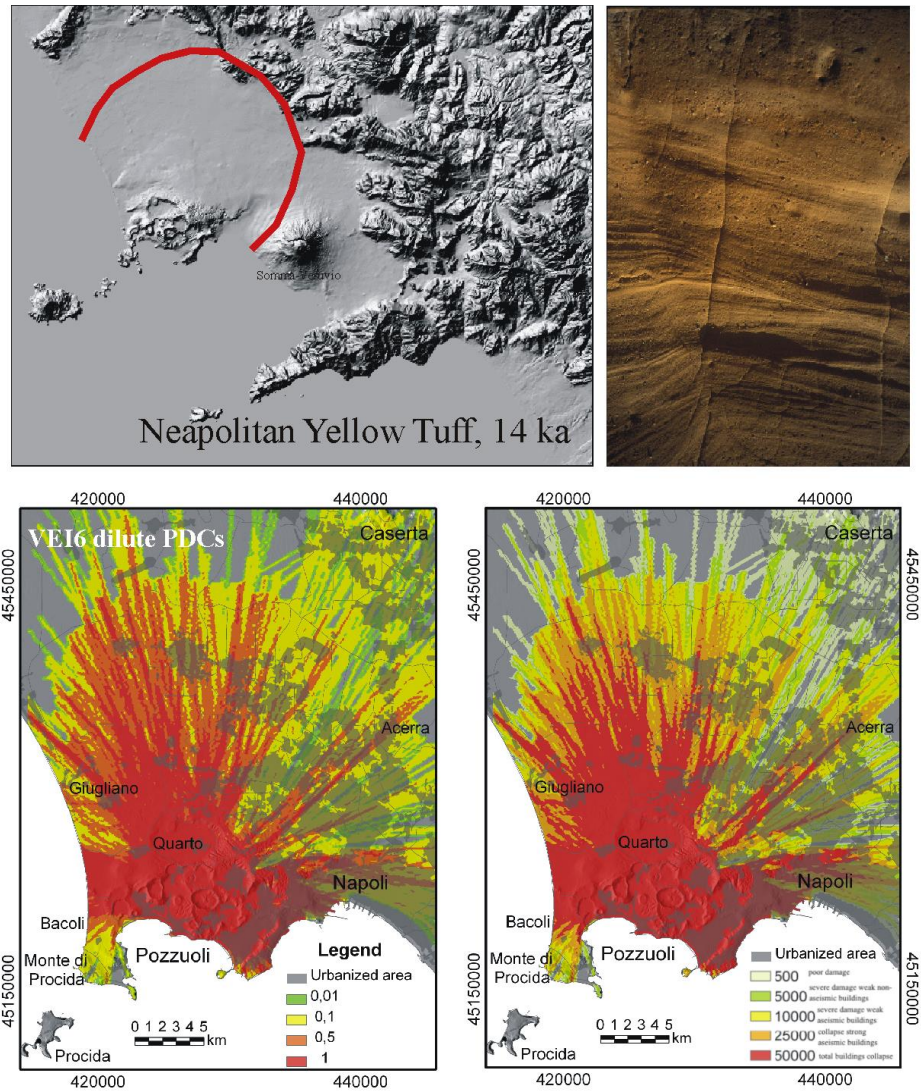

*Figure 6. Model results for VEI 6 eruptions at Campi Flegrei. A) Maximum limit (red line) reached by the PDCs during the ~14.9 ka Neapolitan Yellow Tuff eruption; B) typical exposure of the Neapolitan Yellow Tuff PDC deposits at Posillipo, along the southeastern caldera rim; C) Hazard map of conditional probability and D) the associated maximum dynamic pressures (expressed in Pa), for VEI 6 PDCs (see text and tables 1, 3 for the PDC characteristics adopded in the simulations). Levels of damage associated to different values of dynamic pressure as in Fig. 2. The urbanization pattern is also shown in C-F.*

Discussion and conclusion

Model simulations allow us to explore the areal patterns of PDCs related to a full range of VEI eruptions potentially occurring at Campi Flegrei (Fig. 2-6), with a variety of initial conditions (i.e., PDC velocity, thickness and vent location) and rheological properties. The above maps may describe either PDC trajectories along specific directions, or the whole sectors potentially affected around each vent. The results point out that PDC dispersal varies widely, depending on the size and character of the PDC and vent location. Numerical simulation of PDC trajectories over a topographic model of Campi Flegrei shows that minor PDCs from VEI=2-3 eruptions are strongly controlled by the rugged topography of the volcanic field and are confined in preexisting small valleys and crater relics within a few km from the vent. PDCs with intermediate mobility, from the VEI=4 scenario, for the most part tend to channelize into the valleys contouring pre-existing crater rims and produce irregular distribution patterns. These PDCs are blocked by major topographic barriers, such as Camaldoli, Vomero-Colli Aminei and the western steep slope of Posillipo hill (200 m a.s.l.). The most mobile PDCs, typical of VEI $\geq$5, may travel even in excess of 30 km all around the vent and overpass topographic barriers up to 400 m high.

The computed dynamic pressure values yield a quantitative estimate of the local PDC impact. For relatively dilute PDCs, since model output refers to the behaviour of the leading PDC and neglects locally derived concentrated PDCs, dynamic pressures even higher than reported in Fig. 2-6 may be expected in places. PDC temperature is another essential factor to be considered in risk assessment. Thermal remnant demagnetization analyses (De Gennaro et al. 1999) point out that also wet surges, derived from phreatomagmatic events with the highest water/magma ratio, may retain temperatures even exceeding 200°C as far as the distal reaches. Thus, lethal conditions can occur in the whole area potentially invaded by PDCs, even if flow velocity and dynamic pressure drop to the survival threshold. Overall, the intra-caldera area is always exposed to very high hazard due to the passage of PDCs with high values of dynamic pressure, while the plains north of Campi Flegrei and the Naples city center are exposed to the effects of PDCs with a probability about one order of magnitude less.

The extreme variability of eruptive and emplacement mechanisms recorded in the past eruptions, and reproduced in our simulations, points out that a future eruption may span over a wide range of phenomena and intensities. Previous Authors (e.g., Orsi et al. 2009) considered the geologic record of the last 5000 yrs as a basis for hazard assessment. Although magma composition and volcano-tectonic setting of Campi Flegrei did not vary significantly in the last ~10 ka following the eruptive break after the Neapolitan Yellow Tuff eruption (e.g., Di Vito et al. 1999, Pappalardo et al. 2002), actually the last 3500 yrs have been characterized by substantial quiescence, interrupted occasionally by the AD 1538 Monte Nuovo eruption, which could be the prelude to a new epoch of eruptive activity of un-inferable intensity, style and space-time location.

In light of the above numerical simulations and the existing highly inhomogeneous urbanization pattern (Fig. 2-6), the interplay of the different eruptive parameters and topography determines the risk in a very complex way. The volcanic history documents, for example, that PDCs from the 4.1 ka Agnano-Monte Spina VEI=5 eruption (de Vita et al. 1999), in spite of moderate volume (total 0.5 km$^3$), affected an unusually wide area due to propagation along a sequence of connected valleys. Even slight changes in vent position and/or PDC mobility may result in drastic changes of the exposed value. For example (Fig. 7), a future, relatively small VEI=3 event sourced within the present-day Agnano plain (the highest probability area according to Selva et al. 2012b) would have a substantially different impact in case of vent opening in the eastern part of the plain or in the western part of it, e.g. the area of Pisciarelli where a sensitive unrest has been recorded in the last ten years.

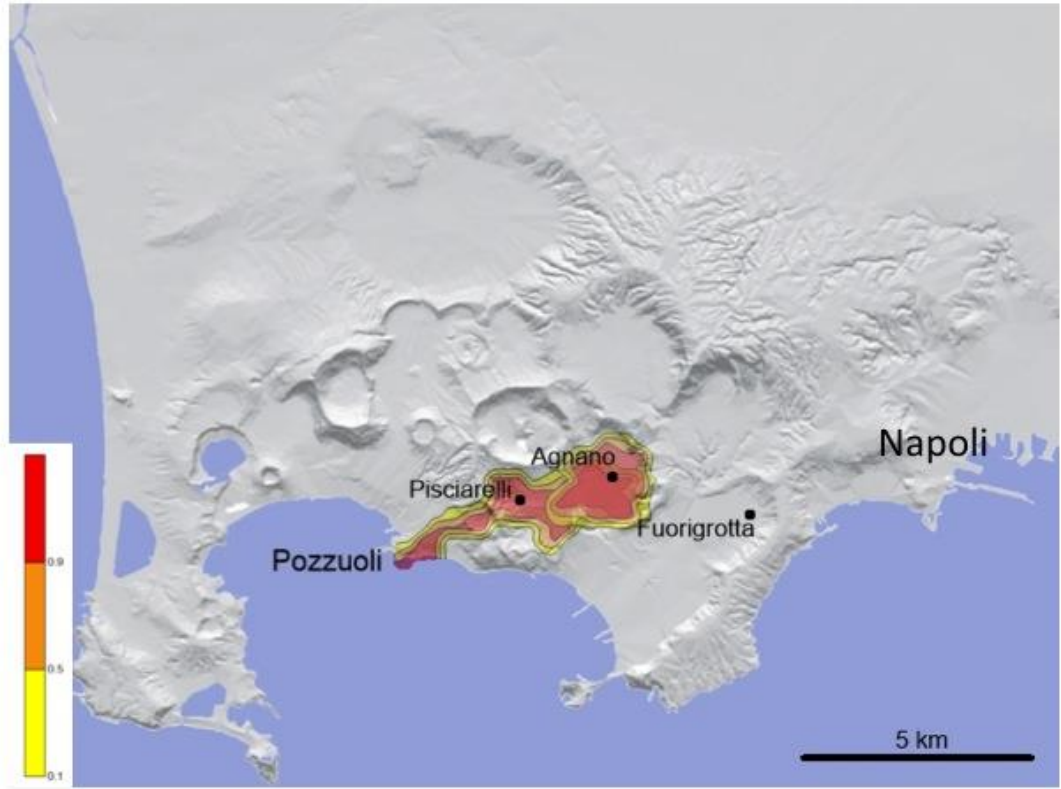

*Figure 7. Examples of the different impact of relatively small (VEI=3) PDC events occurred from two different vents located in contiguous areas (i.e., eastern Agnano plain and Pisciarelli). Left bar shows the conditional probability values. PDC input data: initial velocity: 30 m/s; flow height: 10 m, viscosity=5 Pa/s; yield strength=0; density=50 kg/m³.*

Even human artifacts may act as an additional controlling factor for PDC mobility. Simulation of vent opening near the eastern caldera margin (i.e., Bagnoli-Fuorigrotta plain; Fig. 8) shows that the Fuorigrotta tunnels that connect the Phlegraean area with the city of Naples may drive a small PDC event directly toward the city center, bypassing the barrier of the caldera wall.

On these grounds, the high sensitivity of the PDC distribution pattern to the eruptive source location, makes any inferences on the latter factor critical for risk mitigation strategies.

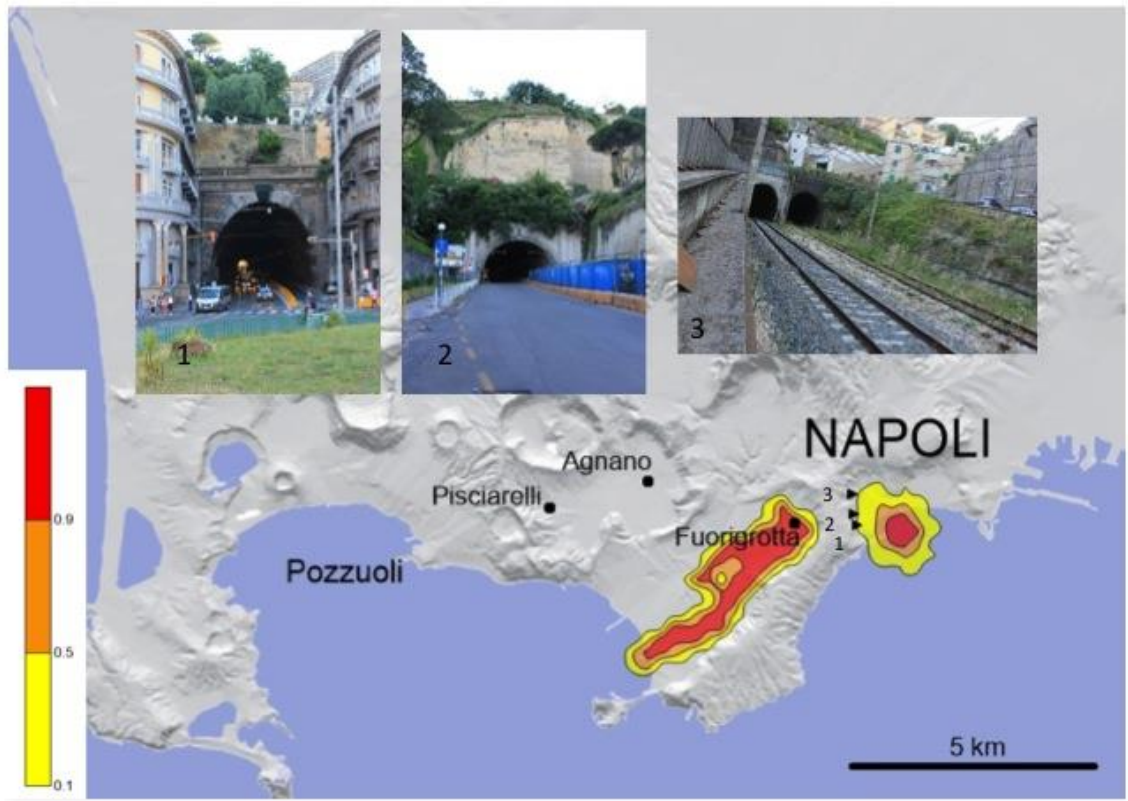

*Figure 8. Example of a small (VEI≤3) PDC event from a vent located near the eastern margin of the caldera (i.e., Bagnoli-Fuorigrotta plain). The presence of road and railway tunnels (1, 2, 3) connecting the Phlegraean area with the city of Naples (Napoli), allows the PDC to bypass the barrier of the caldera wall and impact the city center. Left bar shows the conditional probability values. PDC input data: initial velocity: 20 m/s; flow height: 5 m, viscosity=5 Pa/s; yield strength=0; density=50 kg/m³.*

The possible location of a future eruptive vent within the zone of maximum ground deformation during bradyseismic crises is uncertain. Woo and Kilburn (2010), based on rock mechanic model, instead suggest that the central part of the caldera affected by maximum ground deformation would be the most resistant zone to dike propagation and eruption, while the most likely area for future vent opening would be a ring zone around the caldera centre.

A recently developed probability map for future vent location (Selva et al. 2012b) does not allow a crucial discrimination among the different zones prone to vent opening: except for a slightly higher probability in the Agnano-San Vito zone, the probabilities elsewhere in the area of interest do not deviate significantly from the background. Indeed, the Campi Flegrei history highlights the opening of new vents in different sectors of the caldera during the three recognized volcanic epochs, and thus the future vent location cannot be predicted on the basis of the past 5 ka. Moreover, it is even uncertain how (whether) vent opening probability is influenced by previous eruption occurrences: i.e., whether a future vent location is to be expected within areas of highest vs. lowest vent density. Given the present level of knowledge, eruptive unrest could take place in the whole caldera area, including the highly urbanized intra-caldera plains (e.g., Fuorigrotta, Soccavo, Pianura, Toiano, San Vito). Also, the possible occurrence of an eruption from multiple vents (Isaia et al. 2009) has to be taken into account.

On these grounds, any restricted choice of the zone of vent opening is somehow arbitrary. Given the strong dependence of PDC propagation on the vent position, hazard maps should account for a representative set of potential source locations and surrounding geomorphic conditions. With respect to emergency strategies, a future eruption could be heralded by focused seismicity and/or ground deformation and other signals related to magma ascent (e.g., as reported for the AD 1538 Monte Nuovo eruption; Di Vito et al. 1987), although the identification of the most likely vent opening area would not be possible if not shortly prior to eruption. Then, a selected set of reliable vent locations and PDC scenarios could refine the area of potential PDC impact.

The present work provides implications on the crucial issue of the extension of the evacuation zone in future emergency planning. Fig. 9 summarizes the combined hazard from PDCs (areas exposed to the passage of PDCs with ≥5 kPa maximum dynamic pressure, corresponding to severe building damage/collapse) and concomitant fallout (areas with at least 10% probability of exposure to critical tephra thickness for roof collapse, after Mastrolorenzo et al., 2008) in the case of a VEI 5 eruption from the Campi Flegrei caldera. In our opinion, this is a conservative scenario for the maximum expected event, since its inferred probability of occurrence (4% conditional probability, Orsi et al. 2009) exceeds the negligible threshold of 1% (suggested by Marzocchi and Woo 2009), whereas the worst-case scenarios of VEI 6-7 events only approach the 1% conditional probability limit. In fact, also considering the unpredictability of the event size on the basis of precursors, there is no scientific basis to support the choice of a VEI <5 reference scenario.

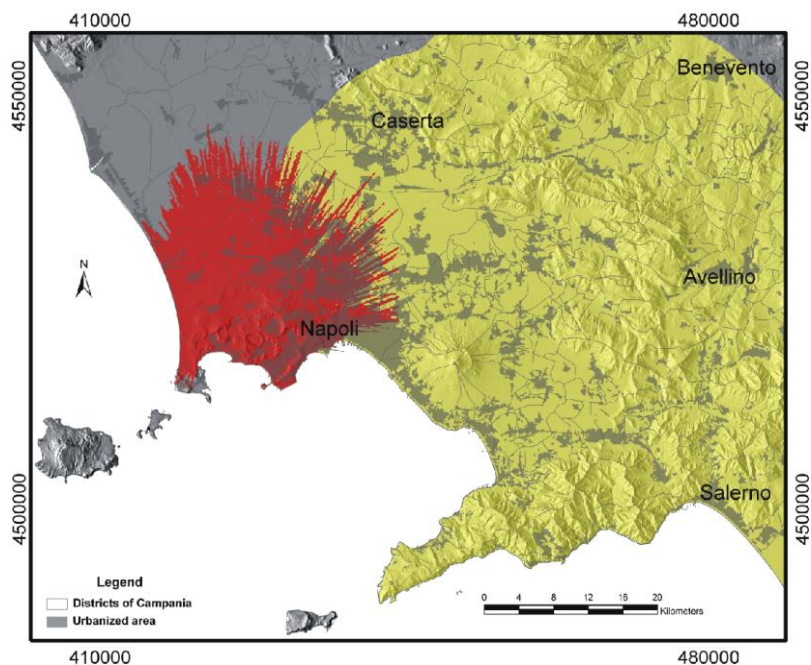

*Figure 9. Combined hazard from PDCs (red: areas exposed to the passage of PDCs with ≥5 kPa maximum dynamic pressure, corresponding to severe building damage/collapse) and concomitant fallout (yellow: areas with at least 10% probability of exposure to critical tephra thickness for roof collapse; after Mastrolorenzo et al., 2008) in case of a VEI 5 eruption that may occur from any vent within the Campi Flegrei caldera, here considered as conservative upper limit scenario.*

Based on the recent activity record (i.e., at least 70 eruptive events occurred in the last 10 ka), an eruptive unrest has an average probability of 0.007 events/yr. The Campi Flegrei caldera has experienced two recent bradyseismic crises, i.e. in 1969-1972 and 1982-84, which could represent a long-term precursor of eruptive unrest (Dvorak and Mastrolorenzo 1991), by analogy with the classic example of the 21 years of unrest (i.e., seismic and ground deformation crises) that preceded the 1994 eruption at Rabaul caldera. In light of the behavior of Rabaul, which, in the last decades has undergone accelerated unrest without eruption, as well as eruption without accelerated unrest until hours beforehand (Robertson and Kilburn, 2012), it can be expected that a future eruption at Campi Flegrei could also occur with only short warning.

Up to now, the extremely risky Campi Flegrei volcanic field lacks an emergency plan. Any reasonable safety strategy should consider a timely evacuation since the early phase of the pre-eruptive alert.

Acknowledgements

We are grateful to Andrea Panizza for his crucial collaboration in the numerical modeling and to Greg Valentine for his comments on an early version of the manuscript. We thank Eliana Bellucci Sessa (Laboratorio di Geomatica e Cartografia, OV-INGV) for plotting the hazard maps in the GIS framework.